\documentclass[11pt,a4paper]{article}
\usepackage[pdftex]{hyperref}
\usepackage{graphicx}
\usepackage{authblk}
\usepackage{amsfonts}
\usepackage{amssymb}
\usepackage{amsmath}
\usepackage{pdfsync}

\begin{document}

\title{Statistical complexity measures as telltale\\ of relevant scales
in emergent dynamics\\ of spatial systems}

\author[1]{A.~Arbona}
\author[1]{C.~Bona}
\author[1]{B.~Mi\~{n}ano}
\author[2]{A.~Plastino}		
\affil[1]{IAC$\,^3$--UIB, Mateu Orfila, Ctra. de Valldemossa km 7.5, 07122, Palma, Spain}
\affil[2]{National University La Plata, Physics Institute (IFLP-CCT-CONICET) C.C. 727, 1900 La Plata, Argentina}

\maketitle

\begin{abstract}
The definition of complexity through Statistical Complexity Measures (SCM) has
recently seen major improvements. Mostly, effort is concentrated in measures
on time series. We propose a SCM definition for spatial dynamical systems.
Our definition is in line with the trend to combine
entropy with measures of structure (such as disequilibrium). We study
the behaviour of our definition against the vectorial noise model of
Collective Motion. From a global perspective, we show how our SCM
is minimal at both the microscale and macroscale, while it reaches a maximum
at the ranges that define the mesoscale in this model. From a local
perspective, the SCM is minimum both in highly ordered and chaotic
areas, while it reaches a maximum at the edges between such areas.
These characteristics suggest this is a good candidate for detecting
the mesoscale of arbitrary dynamical systems as well as regions where
the complexity is maximal in such systems.
\end{abstract}


\maketitle
\newpage

\section{Introduction}

The definition of complexity has recently seen major improvements.
It has been acknowledged for a long
time that simply accounting for information (Shannon or Fisher information,
for instance) does not
fully grasp the notion of complexity, since a perfect chaos maximizes
information but it is actually
not much more complex than perfect order. As Crutchfield noted in 1994,
\textit{Physics does have the tools for detecting and measuring complete order — equilibria
and fixed point or periodic behavior — and ideal randomness — via temperature and thermodynamic
entropy or, in dynamical contexts, via the Shannon entropy rate and Kolmogorov complexity.
What is still needed, though, is a definition of structure and way to detect and to measure it.}~\cite{Crutchfield1994}.

Seth Lloyd counted as many as 40 ways to define complexity, none
of them being completely satisfactory. A major breakthrough came
from the definition proposed by Lopez-Ruiz, Mancini and Calbet
(LMC)~\cite{LopezRuiz1995}. Although not without
problems~\cite{Feldman1998, Kowalsky2011}, LMC's complexity
clearly separated and quantified the contributions of entropy and
structure. LMC measured structure thorough
\textit{disequilibrium}. Building on this proposal, Kowalski et
al.~\cite{Kowalsky2011} refined the definition of disequilibrium.

One should note that while entropy is a general concept that can be
applied across a wide
range of model families, this is not the case with measures of structure. With structure
one needs to know what to seek. Most efforts to define disequilibrium focus on time
series.

We are interested, though, in a statistical complexity measure
(SCM) for models with spatial dimensions. This includes dynamical
PDE-based models, such as Navier-Stokes, and Agent-Based Models
(ABM), such as Collective Motion~\cite{Vicsek2012}. Here we use
the vocable \textit{dynamical} because the  structures we are
interested in are easily recognized (at least visually) by
studying velocity fields. Other models of interest are static
(i.e., not characterized by its velocity field, but rather from
scalar quantities as density or spin). Examples of these models
include PDE-based models such as Reaction-Diffusion, and Cellular
Automata (e.g. Ising models).

Our hypothesis is that, for these systems, a good candidate for
capturing the structural component in the definition of complexity
is a correlation (of the velocity field in the dynamical cases,
and of density or other scalar fields in the static ones). That is

\begin{equation}
C(s(\hat{x},t)) = H(s(\hat{x},t)) \; D(s(\hat{x},t)),
\label{SCM_definition}
\end{equation}
where $H$ stands for entropy (Shannon's, Fisher's, or
Kullback-Leibler's), and $D$ is a correlation. $C(s(\hat{x},t))$
represents the local statistical complexity measure, and
$s(\hat{x},t)$ is the state of the system at time $t$ in position
$\hat{x}$, characterised by some scalar or vector field.

A global SCM is recovered by integration over all the simulation domain $\Omega$:

\begin{equation}
C(s(t)) = \int_{\Omega} d \hat{x} \; C(s(\hat{x},t)).
\end{equation}
$C(s(t))$ is an extensive property. An intensive property is derived by defining

\begin{equation}
C_I(s(t)) = \frac{C(s(t))}{\int_{\Omega} d \hat{x}},
\label{eq.complexity}
\end{equation}
which is simply the average complexity field.

It is also important to acknowledge that the perception of complexity
is deeply imbricated with the scale of measurement. Therefore,
we should aim at measuring $C$ at different scales.
The idea of studying complexity
as a function of scale is not new, as represented for instance
in the concepts of \textit{complexity profile}
\cite{Bar-Yam2002, Bar-Yam2004} and \textit{d-diameter complexity}
\cite{Chaitin1979}. In comparison with these generic frameworks, we
aim at exploring a definition that disentangles the contributions
from entropy and structure and directly exploits the characteristics
of the models under study (velocity fields and density fields in particular)
as a proxy for structure.

Density fields (specifically chemical
concentrations) as a proxy for \textit{structural information}
were studied in the case of reaction-diffusion models
\cite{Lindgren2004}. This work analysed the behaviour of
structural information as both a spatial field and across scales.
We expect this approach to be also fruitful when structure is characterised
by velocity fields rather than density fields, and further enlightened
by separating the contributions of noise and structure.

We aim to study the validity of this hypothesis by  measuring
$C(s(t))$ in numerical simulations of different models.
 In this context we define i)
the microscale, as the size of the simulation mesh (small enough
to represent the microscopical regime characterized by chaotic
dynamics), ii) the mesoscale, as the scale at which a maximum of
complexity is observed (typically characterised by the formation
of turbulence, vortices, clusters, bands, flocks, etc.), and iii) the
macroscale, large enough to reach the hydrodynamical (or
equivalent) limit, typically characterized by different phases and
phase transitions (ordered or chaotic, but not complex).

Representing these three scales in a single simulation is
notoriously difficult in some cases (for instance, a Navier-Stokes
scenario). The microscale and macroscale are separated by many
orders of magnitude. In this case, the strategy should include
different simulations for the distinct scales, adapting the
pertinent equations to each scale and dynamics.

We find a simpler prospect in Collective Motion. Here the dynamics
are characterized by three scales within a few orders of
magnitude, and it is then amenable to a single simulation
encompassing  them all. In this preliminary  study we will focus
attention then  in Collective Motion.

\section{Statistical complexity measures}

Initial value problems with spatial dimensions belong
fundamentally to two big families: mesh-based and meshless. The
former include lattice discretizations of continuous PDE-based
problems, such as Finite Volume Methods or Finite Difference
Methods. It includes, as well, problems directly defined on a
lattice, such as Cellular Automata (Ising models for instance).
Meshless models include discretizations of PDE-based problems that
use particle discretizations (such as Smoothed Particle
Hydrodynamics), as well as problems directly defined on agents or
particles, such as Collective Motion.

In order to define a SCM for such Initial Value Problems, we will
rely on meshes of different cell sizes, defined on the problem's domain
$\Omega$. In the case of mesh-based models, the actual lattice
defining the simulation problem will correspond to the finest SCM
mesh, while coarser SCM meshes for larger scales are trivially
obtained by coarse-graining the finest SCM mesh at various scales.

In the case of meshless models, we define the cell size of the finest SCM mesh
as the typical separation of two agents. The value of a certain field in
a cell of this mesh is obtained by averaging the corresponding fields
of the agents falling into such cell. Coarser meshes are also obtained
by averaging over larger cells.

Let us see how to define a SCM, in this framework, for the case of
an Agent-Based Model characterized by a velocity field (this is
the family of models to which Collective Motion belongs to). That
is, we consider $s=\{\hat{v}\}$. The same procedure is feasible
with other fields (density, spin, and so on). One needs to
properly characterize the fields that, in each case,  better
define the structures representing complexity.

Thus, we consider an agent in an ABM to be characterized, as a
minimum, by its position and velocity. In order to study
complexity at different scales we need to define meshes of various
sizes. As previously stated, each cell in these meshes is
characterized by the same fields as those of the ABM model, with
values computed by averaging the value of the corresponding fields
for all agents falling into the cell. For instance,

\begin{equation}
\hat{v}_i = 1/M \sum_m \hat{v}_m,
\end{equation}
where $\hat{v}_i$ is the velocity in cell $i$, $M$ is the number
of agents in the cell, and $m$ runs over all such agents. If
$M=0$, then $\hat{v}_i=0$.

Following~\cite{Vicsek2012}, and re-scaling so we get a positive
value, we define the velocity correlation as

\begin{equation}
D^l(\hat{v}_i) = \frac{1}{N} \sum_{j \sim i} \frac{ \hat{v}_i \hat{v}_j}
{\hat{v}^2_i + \hat{v}^2_j} + \frac{1}{2},
\end{equation}
where $\hat{v}$ is the velocity vector and the product is scalar.
$i$ defines a particular cell at scale $l$, and $j$ indexes its $N$
immediate neighbours (8 in 2D). For practical purposes
we can only study the cases $\delta x \le l \le L/4$, where $\delta x$ is the
finest mesh size and $L$ is the size of the simulation domain.

This provides us with a local measure of correlation, and through~(\ref{SCM_definition})
a local measure of complexity (a complexity field). Afterwards,
we can integrate this field over all the spatial mesh in order to obtain the global
complexity measure $C^l(\hat{v}(t))$.

In this environment we also compute $H^l(\hat{v}_i)$ as a local
entropy field. We adopt the strategy defined by Xu et al~\cite{Xu2010}.
A first step is to determine a procedure to obtain
a Probability Density Function (PDF). To do so we define a
partitioning of the cell velocity vectors depending on their polar
angle $\theta$, $0 \le \theta \le 2 \pi$, into a number of bins
$x_k$, $k=1...n$ (for 2D systems we set $n=8$).
We continue by estimating the probability for
the vectors in the neighbourhood of cell $i$ as falling into each
bin:

\begin{equation}
p^{l}_{i}(x_k) = {{N(\theta_{j \sim i} \in x_k)} \over{N}}.
\end{equation}

In these circumstances, the  Shannon entropy becomes

\begin{equation}
H^l(\hat{v}_i) = \sum_k p^{l}_{i}(x_k) \; log_2 \; p^{l}_{i}(x_k).
\end{equation}
Once we have defined a local correlation field and a local entropy
field,  we can obtain $C^l(\hat{v}(t))$ as a field for any spatial
position, at any time, and for any (available) scale.

\section{Objectives}

\begin{figure*}
\begin{center}
\includegraphics[width=12cm]{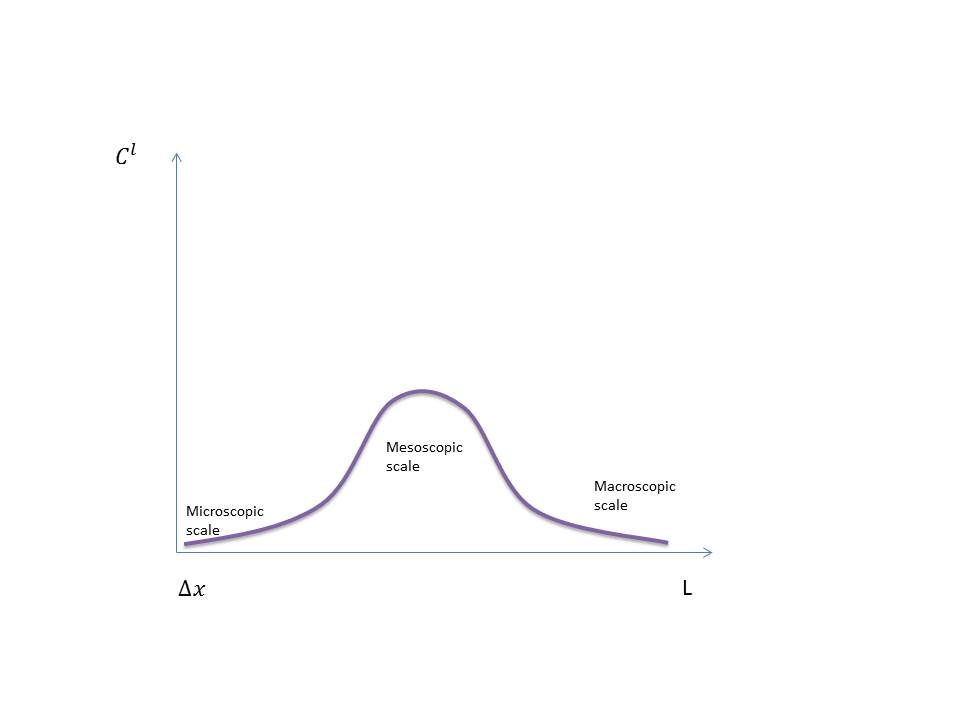}
\caption{Expected complexity versus scale for a model
exhibiting chaotic dynamics at microscale, emergence of
structure at mesoscale, and tendency to isotropy at
macroscale. \label{complexityVersusScales}}
\end{center}
\end{figure*}

We already discussed in the introduction that Collective Motion is
easier to study than more complex models such as Navier-Stokes. A
first step may be to study the shape of $C^l$ for collective
motion, that is, how complexity varies across scales, and thus
confirm the hypothesis that the curve will look like the one
represented in Figure~\ref{complexityVersusScales}.

\begin{figure*}
\begin{center}
\includegraphics[width=8cm]{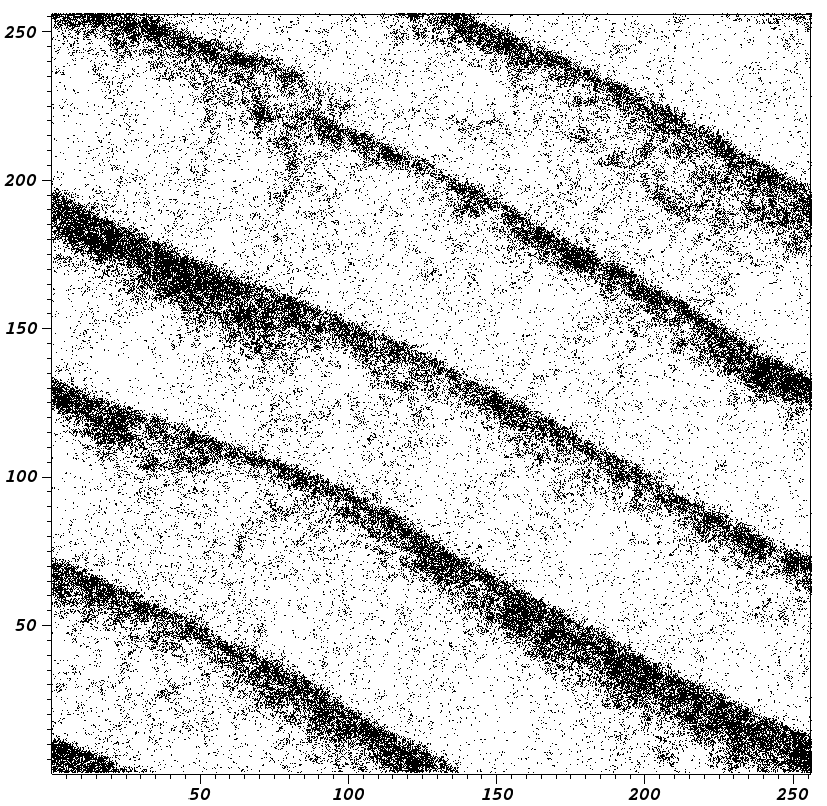}
\caption{Agent distribution in space for a 131.072 particle simulation of
a collective motion model. The model implemented is the vectorial noise model
proposed by Gregoire and Chat\'e~\cite{Gregoire2004},
with domain size $L=256$, density $\rho=2$,
noise level $\eta=0.611$, speed $v_0=0.5$, and time step $\Delta t = 1$.
In this snapshot the system has evolved to a stationary state
characterized by high-density bands travelling in a certain direction.
\label{CM2D_full}}
\end{center}
\end{figure*}

\begin{figure*}
\begin{center}
\includegraphics[width=8cm]{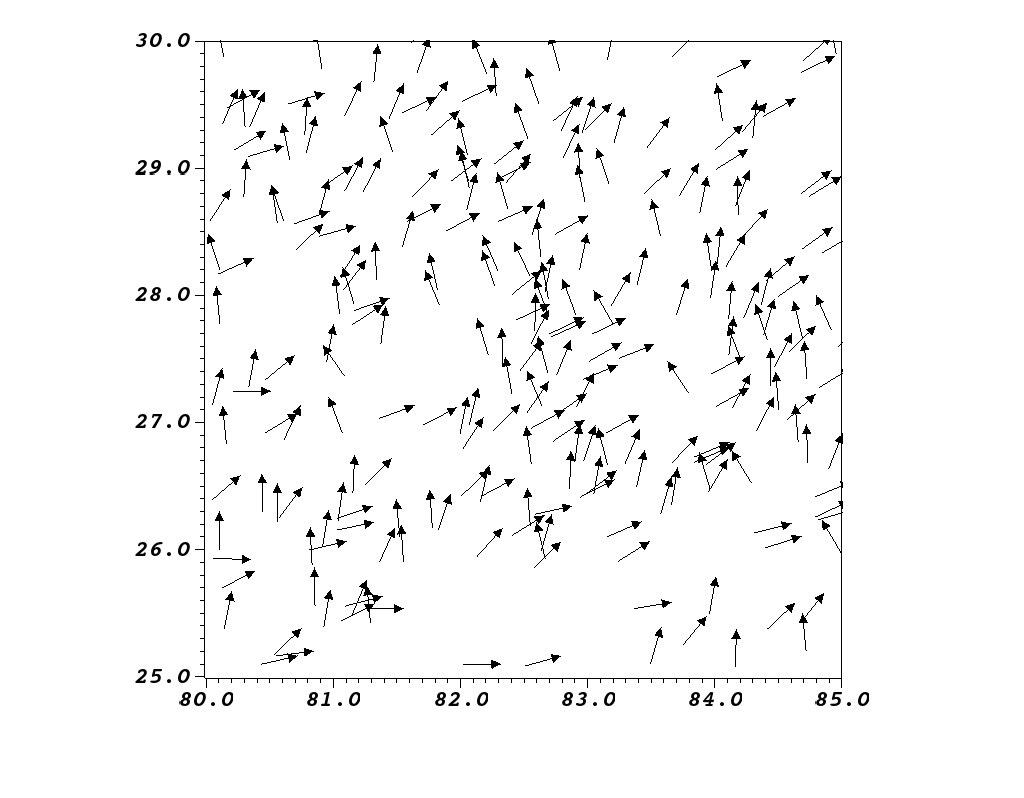}
\includegraphics[width=8cm]{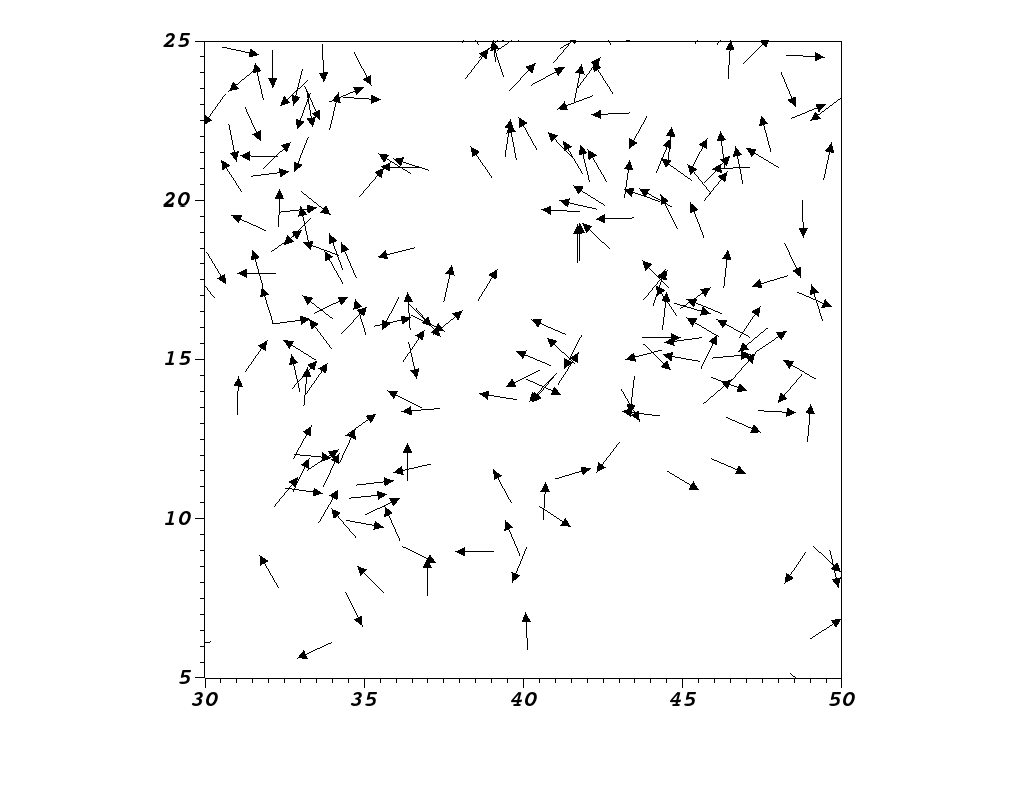}
\caption{Zoom on two small regions of the same simulation represented in
Figure~\ref{CM2D_full}. Arrows correspond to the velocity of each agent. The top figure
zooms into a band; notice the preferred directions North and Northeast.
The bottom figure zooms into an inter-band space, and shows a chaotic pattern of
velocities. \label{CM2D_zoom}}
\end{center}
\end{figure*}

The hypothesis is based on the visual insight one acquires when
looking at figures~\ref{CM2D_full} and~\ref{CM2D_zoom}. These
snapshots are taken from an agent simulation of a Collective
Motion model. Zooming into the velocity field (Figure~\ref{CM2D_zoom})
we see how the chaotic component rules the
dynamics. However, at a larger scale~(Figure \ref{CM2D_full}) the
dynamics is characterized by clustering at a mesoscale, while
isotropy emerges at larger scales (if we adequately zoom out).

\section{Results}
\begin{figure*}
\begin{center}
\includegraphics[width=8cm]{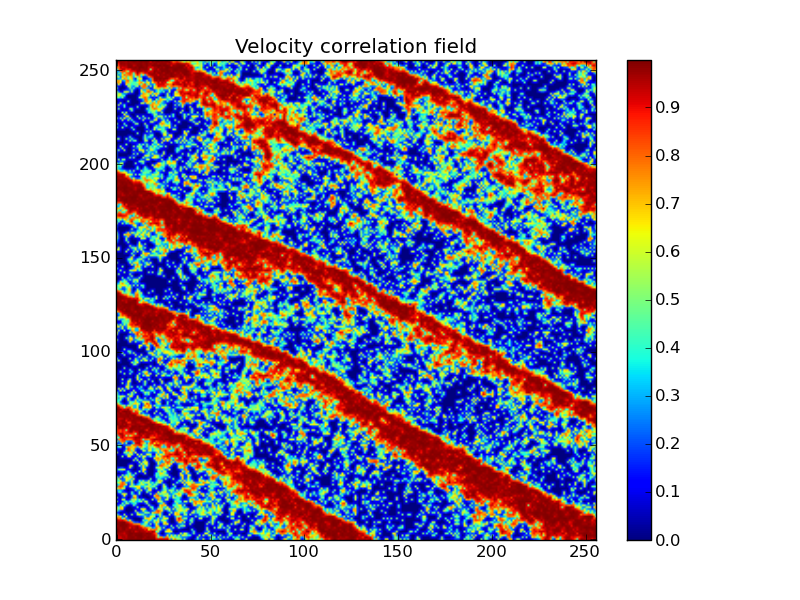}
\includegraphics[width=8cm]{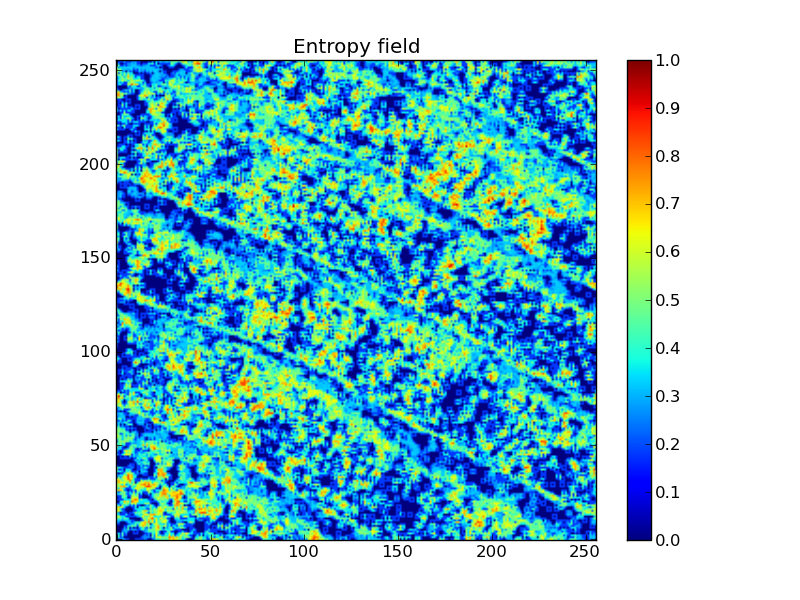}
\includegraphics[width=8cm]{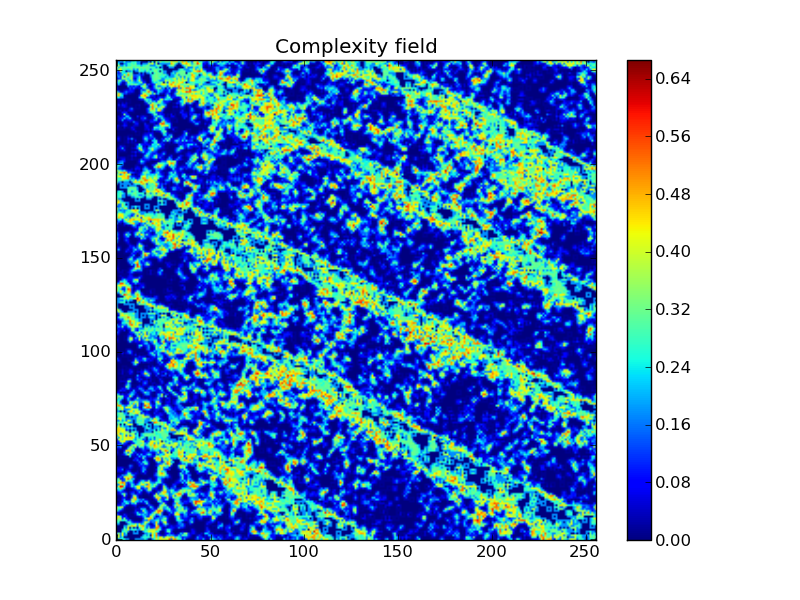}
\caption{Velocity correlation, entropy, and complexity fields for the
snapshot corresponding to Figure \ref{CM2D_full}.
\label{2D_fields}}
\end{center}
\end{figure*}

We have computed the normalised entropy, velocity correlation and
complexity fields for the snapshot considered in
Figure~\ref{CM2D_full}. In Figure~\ref{2D_fields} we show the
results for the finer scale ($\Delta x$, taken as the typical
inter-agent distance). Notice the complementary pattern of
velocity correlation and Shannon entropy. The former is maximal
inside the bands, corresponding with a relatively homogeneous
velocity field. The latter is maximal within the inter-band
regions, particularly in zones when the velocity field is more
disordered. By multiplying these fields, the complexity fields
highlights the areas with a balance between chaos and order,
which correspond with certain locations of the bands (typically
on the band edges, where they make a transition to the turbulent
inter-band regions). Therefore, in general neither the innermost
parts of the bands (highly ordered) nor the inter-band regions
(highly disordered) show the maximum complexity, but rather the
areas that represent a transition between them.

\begin{figure*}
\begin{center}
\includegraphics[width=8cm]{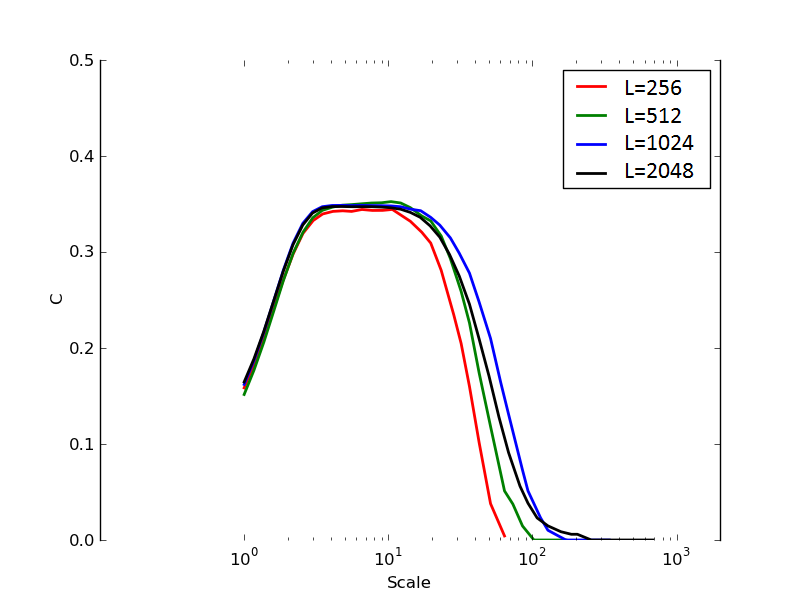}
\caption{The average complexity versus scale of a stationary solution for the
Gregoire and Chat\'e vectorial noise model. The different lines
correspond to 4 different domain sizes: $L=256$, $L=512$, $L=1024$, and $L=2048$.
\label{c_vs_scale}}
\end{center}
\end{figure*}

We have also analysed how complexities changes with scale. To discard effects
produced by the domain size we have run 4 simulations of sizes $256\, x\, 256$,
$512\, x\, 512$, $1024\, x\, 1024$, and $2048\, x\, 2048$. We have taken
snapshots every 500 time steps from $t=20.000$ (where the system is already
in a stationary) to $t=30.000$. For each snapshot and each scale we have computed
the complexity field. We represent in Figure~\ref{c_vs_scale} the average
complexity (averaging both over space within a snapshot and over the set
of snapshots) versus scale. From the comparison among the four plots we
see that the domain size is not affecting the structure of the stationary
state.

The complexity is maximum in the range $3\Delta x - 20\Delta x$, approximately.
The typical band width is about $20\Delta x$, while the transition
zone between bands and inter-band regions has a width of a few $\Delta x$. Thus, the
limits of the mesoscopic scale in this system and the range where complexity
as defined in~(\ref{eq.complexity}) is maximal are barely the same.

\section{Conclusions}
Neither measures of entropy nor measures of structure, taken separately, account
for the accepted notion of complexity. Following the now quite established
trend of combining both quantities into definitions of statistical complexity measures,
we have explored a definition of complexity for spatial dynamical systems.
Our hypothesis is that Shannon entropy is a good proxy for the component of
complexity contributed by disorder, while correlation (velocity correlation
in particular) accounts well for structure.

We have tested our ansatz against the vectorial noise model of
Collective Motion. This is a system with a mesoscale characterised, in
the stationary limit, by travelling ordered bands separated by chaotic
inter-band regions. We have shown that the proposed definition of complexity
field is low both in the most ordered regions, this is the inner part of
the bands, and the most disordered regions (the inter-band regions). The
complexity field is maximum in the band edges, when the ordered dynamics
of the bands meets the chaotic environment of the inter-band regions. This is
precisely the behaviour we expect in a SCM.

We have also studied the average complexity which characterises the
different scales in this system, and found that the complexity is minimum
at both the microscale and the macroscale, a property we were expecting from
a reasonable definition of complexity. For the typical scales that characterize
the mesoscale, we find that the complexity is maximal, in accordance with
our hypothesis.

With these considerations, we conclude that the proposed definition of
statistical complexity measure for a spatial dynamical system is quite
reasonable, and it is a candidate measure for detecting the mesoscale of
arbitrary dynamical systems as well as regions where complexity is
maximum. As a follow-up, we will consider the behaviour of the
proposed definition of complexity for additional spatial dynamical systems.
We will also extend our study to systems characterised by other variables
other than velocity, such as spin or density.

\section{Acknowledgements}
We thank Miquel Trias (Google) for initial feedback on the definition of a SCM for
spatial fields. We thank Joan Mass\'o (University of the Balearic Islands)
for useful discussion on the domain size analysis. We thank Rick Quax
(University of Amsterdam) for illuminating references that improve the
contextualization of our work. The research leading to these results
has received funding from the European
Union Seventh Framework Programme (FP7/2007-2013) under grant agreement no 317534
(the Sophocles project).
\bibliographystyle{plain}
\bibliography{complexity}

\end{document}